\documentclass[prd,twocolumn,floatfix,amsmath,nofootinbib,amssymb,floatfix]{revtex4}
\usepackage{longtable,lscape}
\usepackage{pdfpages}
\usepackage{txfonts}
\usepackage{overpic}
\usepackage
{amssymb}
\usepackage{makecell}
\usepackage{indentfirst}
\usepackage{slashed}  
\usepackage{cases}
\usepackage{multirow}
\usepackage{threeparttable}
\usepackage{enumerate}
\usepackage{subfigure}
\usepackage{graphicx,color,dcolumn,booktabs,bm}
\usepackage{ulem}
\usepackage[colorlinks,
            citecolor=blue,
            anchorcolor=red,
            menucolor=red,
            linkcolor=red,
            filecolor=red,
            runcolor=red,
            urlcolor=blue,
            frenchlinks=red]{hyperref}

\graphicspath{{figures/}}

\begin{document}

\title{Double-bottom centrifugal-barrier molecules dancing with four quarks}

\author{Zhi-Peng Wang$^{1,2}$}
\email{wangzhp2020@lzu.edu.cn}
\author{Fu-Lai Wang$^{1,2,3,4,5}$}
\email{wangfl2016@lzu.edu.cn}
\author{Guang-Juan Wang$^{6}$}
\email{wgj@post.kek.jp}
\author{Xiang Liu$^{
1,2,3,4,5}$}
\email{xiangliu@lzu.edu.cn}
\affiliation{$^1$School of Physical Science and Technology, Lanzhou University, Lanzhou 730000, China\\
$^2$Research Center for Hadron and CSR Physics, Lanzhou University and Institute of Modern Physics of CAS, Lanzhou 730000, China\\
$^3$Key Laboratory of Quantum Theory and Applications of MoE, Lanzhou University,
Lanzhou 730000, China\\
$^4$Lanzhou Center for Theoretical Physics, Key Laboratory of Theoretical Physics of Gansu Province, Lanzhou University, Lanzhou 730000, China\\
$^5$MoE Frontiers Science Center for Rare Isotopes, Lanzhou University, Lanzhou 730000, China\\
$^6$KEK Theory Center, Institute of Particle and Nuclear Studies (IPNS), High Energy Accelerator Research
Organization (KEK), 1-1 Oho, Tsukuba, Ibaraki 305-0801, Japan}

\begin{abstract}
Motivated by the recent Belle II measurement of the open-bottom cross section, we perform a systematic study of the double-bottom molecular tetraquark spectrum, including both $S$-wave and $P$-wave configurations. Using the one-boson-exchange potential and the complex scaling method, we investigate the $B^{(*)}\bar{B}^*$ and $B^{(*)}B^*$ systems and predict several bound states and resonances. Our analysis reveals a rich spectrum, including $B\bar{B}^*$ ($0(1^{++})$, $0(0^{-+})$), $B^*\bar{B}^*$ ($0(2^{++})$, $0(0^{-+})$, $0(1^{--})$), $BB^*$ ($0(1^+)$), and $B^*B^*$ ($0(1^+)$) bound states, as well as $B\bar{B}^*$ ($0(1^{--})$, $0(2^{--})$), $B^*\bar{B}^*$ ($0(3^{--})$), $BB^*$ ($0(0^-)$), and $B^*B^*$ ($0(1^-)$, $0(2^-)$) resonances.  Our results provide theoretical support for the existence of double-bottom tetraquarks and deliver key benchmarks for future searches at LHCb, Belle II, and other experiments.
\end{abstract}

\maketitle

\section{Introduction}The investigation of exotic hadrons with minimal quark contents beyond the conventional $q\bar q$  or $qqq$ pictures~\cite{GellMann:1964nj,Zweig:1981pd} has become a frontier in particle physics, offering insight into the nonperturbative regime of strong interactions. Over the past two decades, the heavy-flavor exotic states has grown rapidly and attracted intense interest following the discovery of many  near-threshold hadronic candidates, such as $X(3872)$ \cite{Belle:2003nnu}, $Z_c(3900)$ \cite{BESIII:2013ris}, $Z_b(10610)$, $Z_b(10650)$ \cite{Belle:2011aa}, and so on. Their proximity to the open-flavor thresholds naturally suggests the hadronic-molecule interpretation and many candidates are reviewed in
Refs.~\cite{Liu:2013waa,Hosaka:2016pey,Chen:2016qju,Richard:2016eis,Lebed:2016hpi,Brambilla:2019esw,Liu:2019zoy,Chen:2022asf,Olsen:2017bmm,Guo:2017jvc,Meng:2022ozq}. Yet their internal structures remain unresolved.  

While most hadronic molecules discussed so far are $S$-wave, there is growing interest in possible resonant structures in higher partial waves. Several studies have shown that the hadron-hadron interactions still remain attractive albeit weaker in higher partial wave and centrifugal barrier arsing from the orbital angular momentum stabilizes resonant states even under weak inter-hadron attraction.  In Ref. \cite{Wang:2023ivd},  we showed that if the renowned $Y(4260)$, $Y(4360)$, and $\psi(4415)$ are identified as the $S$-wave molecules,  there will be $13$ $P$-wave resonances and $4$ $S$-wave bound counterparts, including $8$ exotic $J^{PC}$ states forbidden to $c\bar c$ mesonic system. 

In Refs. \cite{Ohkoda:2011vj,Ohkoda:2012hv}, Ohkoda \textit{et al.} performed a systematic investigation of the doubly heavy molecular tetraquark states with total angular momentum $J\le2$, adopting the $\pi/\rho/\omega$ meson exchange interactions. Their results revealed a rich spectrum dominated by $P$-wave resonances. 
Ref.~\cite{Wang:2024ukc} suggested that the double pole structures of the manifestly exotic tetraquark state $T_{cs1}(2900)$ \cite{LHCb:2020pxc} can be interpreted as the $
P$-wave $\bar{D}^{*}K^*$ dimeson resonances, and recent work has proposed that the $G(3900)$ structure \cite{BaBar:2006qlj} can be assigned to the $P$-wave $D\bar{D}^*$ resonance with $I(J^{PC})=0(1^{--})$ \cite{Lin:2024qcq}. Furthermore, the molecular tetraquark candidates with $P$-wave excitations have been systematically explored in both the bottom and charm quark sectors \cite{Sakai:2025djx,Lu:2025zae,Chen:2025gxe}, and possible exotic baryons with $P$-wave hadronic molecular configurations have also been predicted \cite{Yu:2021lmb}.
These findings underscore how $P$-wave dynamics can enrich the spectrum of the exotic hadrons.
Taken together, these studies suggest that {\it Centrifugal-Barrier Molecules} may constitute a new group of the hadron spectroscopy.

\begin{figure}[htbp]
    \centering
    \includegraphics[width=0.48\textwidth]{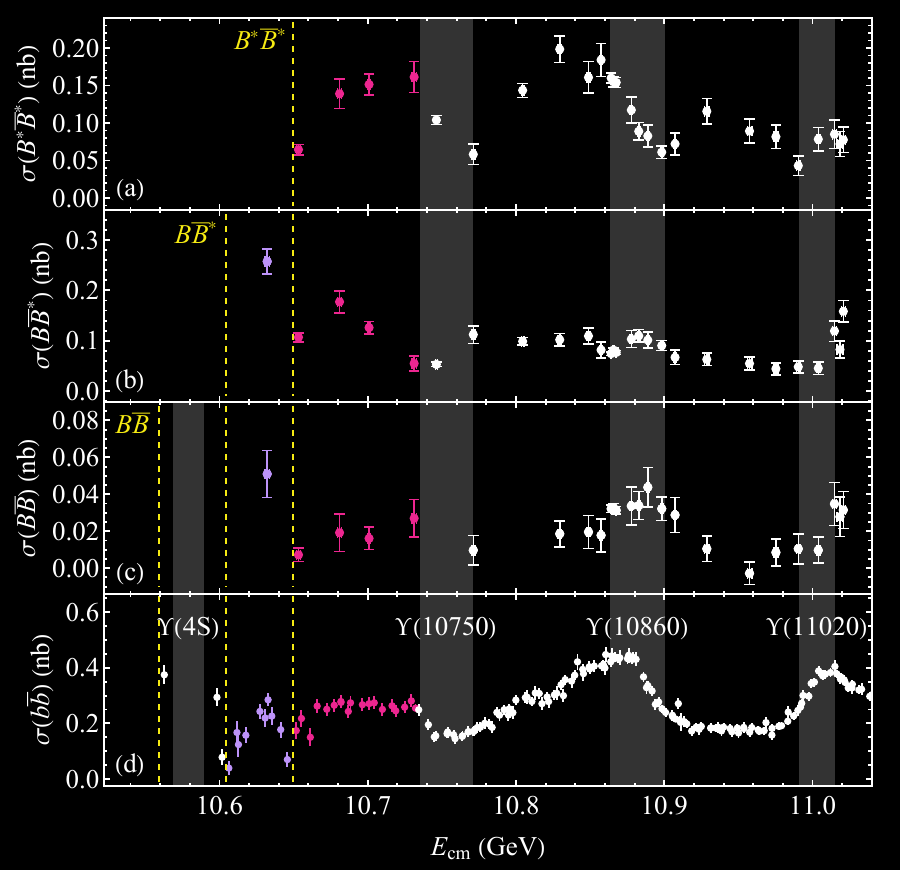}
    \caption{(Color online.) The various open bottom cross section. The exclusive cross section of the $e^+e^-\to B^*\bar{B}^*$, $e^+e^-\to B\bar{B}^*$, and $e^+e^-\to B\bar{B}$ processes is taken from Belle~\cite{Belle:2021lzm} and Belle II~\cite{Belle-II:2024niz}. The inclusive $e^+e^-\to b\bar{b}$ cross section is taken from Ref.~\cite{Dong:2020tdw}.}
    \label{fig:data}
\end{figure}

The bottom quark sector serves as an ideal laboratory for exploring exotic hadronic states. In the past decades, extensive theoretical studies have explored potential $B^{(*)}\bar{B}^{(*)}$ bound states \cite{Sun:2011uh,Zhao:2014gqa,Chen:2015ata,He:2014nya,Guo:2013sya,Baru:2019xnh}, with several works predicting the existence of the $S$-wave $B^{(*)}\bar{B}^{(*)}$ bound states \cite{Liu:2013waa,Hosaka:2016pey,Chen:2016qju,Richard:2016eis,Lebed:2016hpi,Brambilla:2019esw,Liu:2019zoy,Chen:2022asf,Olsen:2017bmm,Guo:2017jvc,Meng:2022ozq}. Very recently, existing data already suggest nontrivial dynamics in the $B^{(*)}\bar{B}^{(*)}$ sector (See Fig.~\ref{fig:data}). Notably, there are unconventional peak and dip structures in the observed line shapes. Specially, the rapid rise of $\sigma(B^*\bar B^*)$ near the threshold suggests the presence of nontrivial dynamics, which may originate from the molecular configurations in the $B^{(*)}\bar{B}^{*}$ systems.

Motivated by these experimental hints, we present an unified complex-scaling analysis of both $S$- and $P$-wave \(B^{(*)}\bar B^{(*)}\) and \(B^{(*)}B^{(*)}\) systems within the one-boson-exchange (OBE) mechanism. To obtain the bound and resonant state solutions, we employ the complex scaling method (CSM). We predict various bound and resonant states. Specifically, two $P$-wave $B\bar{B}^*$ and $B^*\bar{B}^*$ molecules with quantum number $I(J^{PC})=0(1^{--})$ are found and may account for the line shape anomalies in $e^+e^-$ annihilation. 
Additionally, we explore the $B^{(*)}B^*$ systems, which are related to the $B^{(*)}\bar{B}^*$ systems through the $G$-parity rule. These states are of ongoing Belle II scans and will be further testable at LHCb Upgrade II. 

Once these states are observed, confirmation of these predictions would establish centrifugal-barrier binding as a new mechanism in heavy spectroscopy, lead to the identification of new bottom-sector $P$-wave molecules, and help to examine the $S$-wave partner states.

\section{The $B^{(*)}\bar{B}^*/B^{(*)}{B}^*$ interactions and the ${\rm CSM}$}
Under the heavy quark spin symmetry, the heavy quark spin doublet $(\bar{B},\bar{B}^*)$ can be described by a superfield $H_a$, i.e., $H_a=\frac{1+\slashed{v}}{2}\left(\bar{B}^{*\mu}_a\gamma_{\mu}-\bar{B}_a\gamma_5\right)$, and its conjugate field is defined as $\overline{H}_a=\gamma^0 H^\dagger_a\gamma^0$. Guided by the interplay of the heavy-quark symmetry, chiral symmetry, and hidden local symmetry, the effective Lagrangians governing interactions between the heavy hadrons and the light scalar, pseudoscalar, and vector mesons can be constructed as \cite{Burdman:1992gh,Wise:1992hn,Yan:1992gz}
\begin{align}\label{Lagrangians}
\mathcal{L}_{\bar{B}^{(*)}\bar{B}^{(*)}}= & \left\langle g_\sigma H_a\sigma\overline{H_a}\right\rangle+\left\langle ig H_b \mathcal{A}\!\!\!\slash_{ba}\gamma_5\overline{H_a}\right\rangle \nonumber \\
&+\left\langle i H_b\left(\beta v^\mu\left(\mathcal{V}_\mu-\rho_\mu\right)+\lambda \sigma^{\mu \nu} F_{\mu \nu}(\rho)\right)_{b a} \overline{H}_a\right\rangle,
\end{align}
where the axial current $\mathcal{A}_\mu$, the vector current $\mathcal{V}_\mu$, the vector meson field $\rho_\mu$, the vector meson strength tensor $F_{\mu\nu}$, the light pseudoscalar meson matrix $\mathbb{P}$, and the light vector meson matrix $\mathbb{V}_\mu$ follow standard definitions, which can be found in Refs.  \cite{Burdman:1992gh,Wise:1992hn,Yan:1992gz}. For numerical calculations, we adopt the following values for the coupling constants \cite{Sun:2012zzd,Li:2012cs,Li:2012ss}, i.e., $g_\sigma=0.76$, $g=0.59$, $f_\pi=132~\rm{MeV}$, $\beta=0.90$, $\lambda=0.56 ~\rm{GeV}^{-1}$, and $g_V=5.83$.

Starting from the leading-order effective Lagrangians in Eq.~\eqref{Lagrangians}, the effective potentials $\mathcal{V}(\bm{q})$ for the $B\bar{B}^*$ and $B^*\bar{B}^*$ systems can be derived from the scattering amplitudes $\mathcal{M}(\bm{q})$ within the Breit approximation \cite{Berestetskii:1982qgu}, i.e.,
\begin{equation}
    \mathcal{V}(\bm{q})=-\frac{\mathcal{M}(\bm{q})}{\prod_i\!\sqrt{2m_{h_i}}}\mathcal{F}^2(q^2,m_E^2) \,,
\end{equation}
where $m_{h_i}$ and $m_E$ denote the masses of hadron $h_i$ and exchanged mesons $E$, respectively. The monopole form factor $\mathcal{F}(q^2,m_E^2)=(m_E^2-\Lambda^2)/(q^2-\Lambda^2)$ is introduced \cite{Tornqvist:1993ng,Tornqvist:1993vu}, where $\Lambda$ represents the cutoff parameter, and $q$ is the four momentum of the exchanged light mesons. In the present work, we adopt a cutoff value of 1 GeV, motivated by the successful description of the deuteron, $P_{c}$, and $T_{cc}(3875)^+$ as the hadronic molecules using similar cutoff values \cite{Urey:1932gik,Yang:2011wz,Chen:2015loa,Li:2012ss}. 
In fact, as we will show, the existence of the obtained states is robust against the variations of the cutoff around 1~GeV, although the precise determination of their resonance parameters requires future experimental results with sufficient accuracy. 
This choice ensures consistency with the established molecular state calculations. Furthermore, according to the $G$-parity rule \cite{Klempt:2002ap,Lin:2024qcq}, the interactions of the $BB^*$ and $B^*B^*$ systems can be related to those of the $B\bar{B}^*$ and $B^*\bar{B}^*$ systems, respectively. This rule enables a unified treatment of both particle-particle and particle-antiparticle interactions within the same framework.

To explore the bound and resonant states for the $B^{(*)}\bar{B}^*$ and $B^{(*)}B^*$ systems, we utilize the CSM \cite{Myo:2014ypa,Myo:2020rni,Moiseyev:1998gjp}, which introduces a transformation operator $U(\theta)$ acting on the radial coordinate ${\bm r}$ and its conjugate momentum ${\bm p}$, i.e.,
\begin{eqnarray*}
U(\theta)\bm{r}U^{-1}(\theta)=\bm{r}e^{i\theta}~~~{\rm and}~~~
U(\theta)\bm{p}U^{-1}(\theta)=\bm{p}e^{-i\theta}.
\end{eqnarray*}
Thus, the transformated Schr\"odinger equation in the momentum space becomes
\begin{align}\label{transformation}
    &\frac{p^2 e^{-2i\theta}}{2\mu}\Psi_{L}^{\theta}(p)+\int \frac{e^{-3i\theta}p^{\prime 2}\mathrm{d} p'}{(2\pi
)^3}V_{LL}^{JS}(p e^{-i\theta},p'e^{-i\theta})\Psi_{L}^{\theta}(p')\nonumber\\
    &=E(\theta)\Psi_{L}^{\theta}(p).
\end{align}
Here, the potential $V_{LL}^{JS}(p,p')$ can be expressed as \cite{Jacob:1959at}
\begin{align}
V_{LL}^{JS}(p,p')&=\sum_{\lambda_1,\lambda_2,\lambda_3,\lambda_4}\mspace{-7mu} U_{\lambda_3\lambda_4}^{J,LS}\int {\rm d}{\Omega_{\bm p\bm p'}}D_{\lambda\lambda'}^J(\Omega_{\bm p\bm p'},0)\nonumber\\
&\quad \times\langle \bm p\lambda_3\lambda_4|\mathcal V|\bm p'\lambda_1\lambda_2\rangle \left(U_{\lambda_1\lambda_2}^{J,LS}\right)^\dagger,
\end{align}
where $\lambda_i$ represents the helicity of the corresponding particle $i$ with  $\lambda=\lambda_1-\lambda_2$ and $\lambda'=\lambda_3-\lambda_4$. $\Omega_{\bm p\bm p'}$ denotes the solid angle between the directions of the momenta $\bm p$ and $\bm p'$, and $D_{\lambda\lambda'}^J(\Omega_{\bm p\bm p'},0)$ represents the Wigner-$D$ function.
$U_{\lambda_1\lambda_2}^{J,LS}$($U_{\lambda_3\lambda_4}^{J,LS}$) is the transformation matrix between $|JMLS\rangle$ and the two-particle spherical-wave helicity basis $|JM\lambda_1\lambda_2\rangle$($|JM\lambda_3\lambda_4\rangle$).

According to the ABC theorem \cite{Aguilar:1971ve,Balslev:1971vb}, the bound and resonant state solutions can be extracted from the eigenvalues $E(\theta)$ of the transformated Schr\"odinger equation. Specifically, the bound states reside on the negative real energy axis and are independent of the scaling angle $\theta$, the resonant states, characterized by complex energies $E=E_{r}-i\Gamma_{r}/2$, emerging and remaining independent of $\theta$ when $\theta$ satisfies the condition ${\rm Tan}^{-1} (\Gamma_{r}/2E_{r})<2\theta$.

\section{The mass spectra of the hidden-bottom and open-bottom molecular tetraquark states}
\begin{figure}[htbp]
    \centering
    \includegraphics[width=0.48\textwidth]{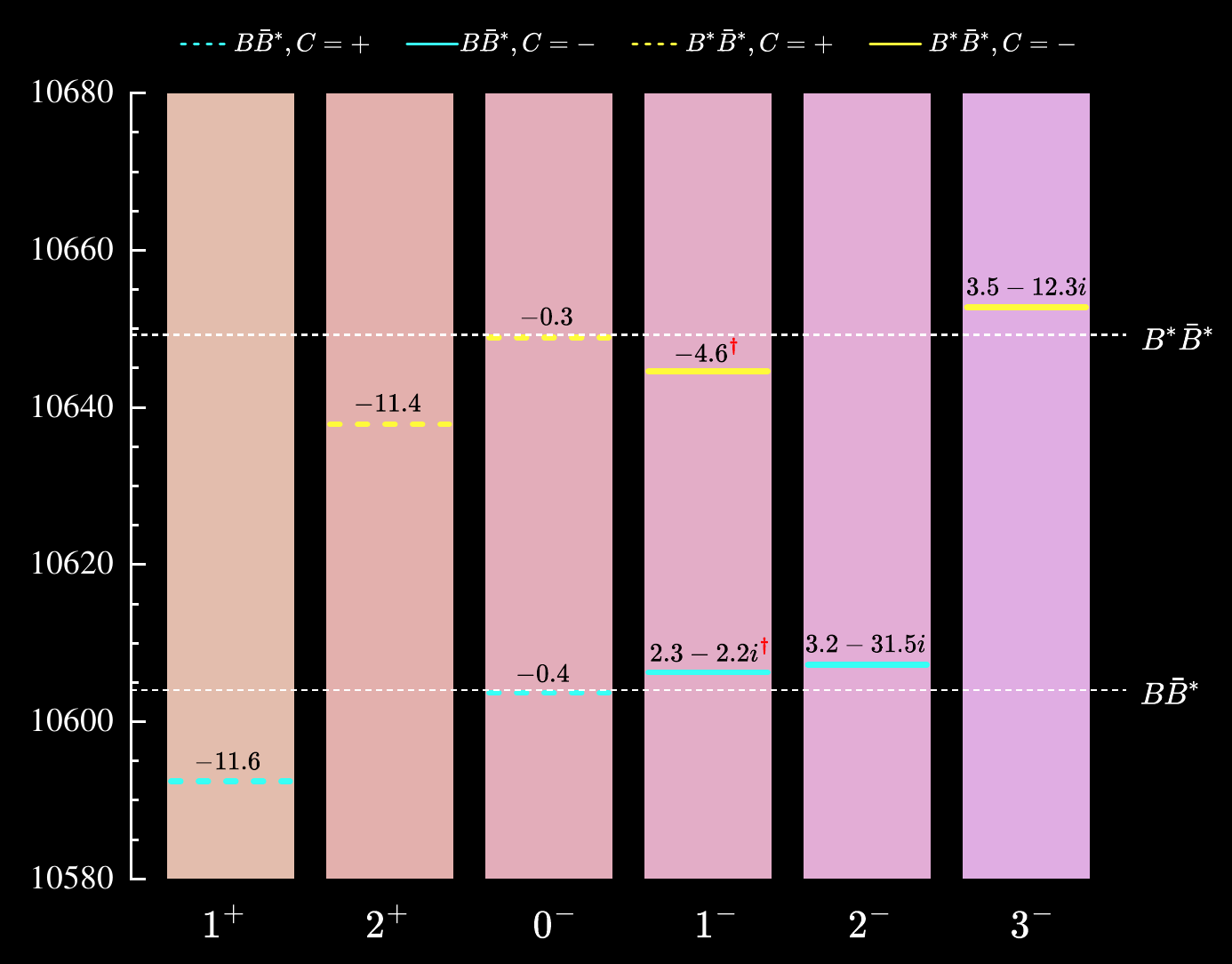}
    \caption{The mass spectra of the $I=0$ molecular states for the $B\bar{B}^*$ and $B^*\bar{B}^*$ systems, including the bound states (denoted by binding energies $E_b$) and the resonances (expressed as $E_r - i\Gamma_r/2$). Here, the cutoff value is fixed at 1 GeV, and the two $1^{--}$ molecules are marked with the symbol $\dagger$.}
    \label{fig:hb_spectrum}
\end{figure}

Employing the OBE mechanism and the CSM, we first calculate the mass spectra of the $B\bar{B}^*$ and $B^*\bar{B}^*$ systems across all channels up to the $P$-wave, and the corresponding mass spectra are illustrated in Fig. \ref{fig:hb_spectrum}. 
For the $B\bar{B}^*$ system, we identify two bound states: one in the $0(1^{++})$ channel and another in the $0(0^{-+})$ channel. Notably, the latter forms a shallow $P$-wave bound state with a binding energy of only a few tenths of an MeV. This state evolves into a near-threshold resonance when the interaction strength is slightly decreases. For the $B^*\bar{B}^*$ system, the bound states with $0(2^{++})$ and $0(0^{-+})$ are obtained as the heavy quark spin partner of the $0(1^{++})$ and $0(0^{-+})$ $B\bar{B}^*$ states, respectively. 
In addition to the $P$-wave bound state in the $0(0^{-+})$ channel, a shallow $P$-wave $B^*\bar{B}^*$ bound state with $0(1^{--})$ is also identified. This state involves two coupled partial waves, $^{1}P_1$ and $^{5}P_1$, with the latter being the dominant component. Although the dominant one-pion-exchange potential in $^{1}P_1$ partial waves is repulsion, the attractive interaction in $^{5}P_1$ channel plays a crucial role in the formation of this bound state, and the off-diagonal attractive interaction between the $^{1}P_1$ and $^{5}P_1$ partial waves contributes to the stabilisation of this state. 

Besides the bound molecules, several resonances are predicted: the $B\bar{B}^*$ states with the quantum numbers $0(1^{--})$ and $0(2^{--})$, and the $B^*\bar{B}^*$ states with $0(3^{--})$. 
All predicted $B\bar{B}^*$ and $B^*\bar{B}^*$ resonant molecules belong to the $P$-wave systems, where the orbital angular momentum plays a critical role in generating these resonant structures \cite{Wang:2023ivd,Lin:2024qcq}. Unlike $S$-wave systems, the $P$-wave centrifugal barrier creates a confining potential necessary for the resonance formation.

In addition, these resonances appear near threshold energies, typically around tens of MeV, and exhibit significantly narrower widths compared to their charm-sector counterparts \cite{Wang:2023ivd,Lin:2024qcq,Ohkoda:2012hv,Lu:2025zae}.
For example, in the $0(1^{--})$ channel, the $D\bar{D}^*$ resonant predicted in Ref. \cite{Lin:2024qcq} has a width of 50 MeV, whereas the $B\bar{B}^*$ resonant has a width of only a few MeV.   
The suppressed widths in the bottom sector originate from the larger reduced mass of the $b$-quark systems, which suppresses both kinetic energy and repulsive centrifugal potential-each inversely proportional to the reduced mass of the system. In the heavy quark limit, this suppression becomes even more pronounced, causing the widths of these resonances to vanish and the poles to transition into the bound states.
For the $B\bar{B}^*$ system with $C=-$, the width of the $J^P=2^-$ resonance is an order of magnitude greater than that of the $1^-$ resonance. 
This difference stems from the weaker attraction in the dominantly one-pion-exchange interaction for the higher-spin state.

\begin{figure}[htbp]
    \centering
    \includegraphics[width=0.48\textwidth]{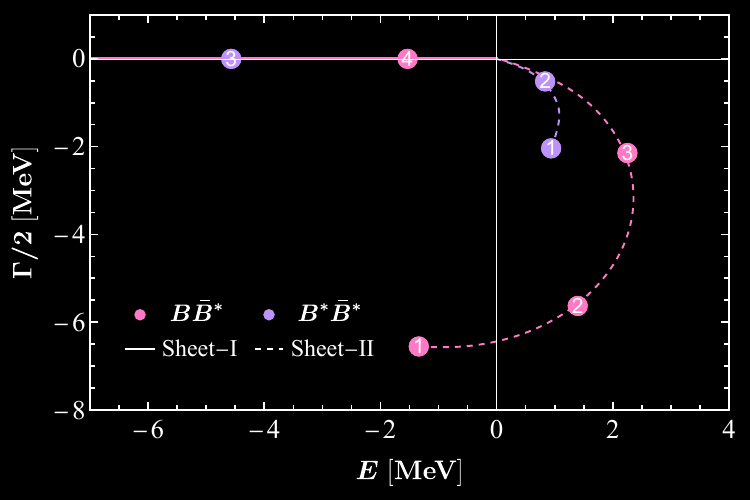}
    \caption{The pole trajectories with the cutoff values correspond to the $B\bar{B}^*$ and $B^*\bar{B}^*$ $0(1^{--})$ molecules. The circled number 1–4 represent the increasing cutoff 0.8–1.1 GeV in order. The solid (dashed) lines represent the pole trajectories in the first (second) Riemann sheets.}
    \label{fig:cutdepen}
\end{figure}

In Fig. \ref{fig:cutdepen}, we present the pole trajectories of the $B\bar{B}^*$ and $B^*\bar{B}^*$ $0(1^{--})$ molecular states as the cutoff value varies from 0.8 to 1.1 GeV. When the cutoff is below 1.0 GeV, both states manifest as near-threshold resonances. As the cutoff further increase, the pole associated with the $B^*\bar{B}^*$ states first moves into the physical Riemann sheet and becomes a bound state, followed subsequently by the $B\bar{B}^*$ states. At a cutoff 1.1 GeV, the $B\bar{B}^*$ and $B^*\bar{B}^*$ states exhibit binding energies of a few and about 20 MeV, respectively. These results indicate that the existence of near-threshold $B^{(*)}\bar{B}^*$ states remains robust under reasonable variations of the cutoff parameter.

The interaction of the $BB^*$ system can be related to those of the $B\bar{B}^*$ system via the $G$-parity rule \cite{Klempt:2002ap,Lin:2024qcq}. Building on this connection, we extend our study to investigate potential bound and resonant states for the $BB^*$ and $B^*B^*$ systems. The corresponding results are summarized in Table \ref{tab:results2}. For the $BB^*$ system, we identify a bound state with $I(J^P)=0(1^+)$, which corresponds to the bottom-sector counterpart of the well-established $T_{cc}(3875)^+$ \cite{LHCb:2021vvq}. Additionally, a resonant state with $I(J^P)=0(0^-)$ is found near the $BB^*$ threshold, characterized by a narrow width. For the $B^*B^*$ system, we predict a bound state with quantum number $I(J^P)=0(1^+)$ and a binding energy of 13.9 MeV, which is consistent with the result from the chiral effective field theory, predicting a binding energy of 12.6 MeV \cite{Wang:2018atz}.  Notably, the $B^*B^*$ configurations with $I(J^P)=0(1^-)$ and $0(2^-)$ exhibit the resonant behaviors. These states represent critical candidates for molecular tetraquarks in the bottomonium sector, completing the spectroscopic landscape alongside their hidden-bottom counterparts.

\begin{table}[!htbp]
    \caption{The bound and resonant states for the $BB^*$ and $B^*B^*$ systems up to the $P$-wave. Here, the units of the energy and half-width are $\rm {MeV}$.\label{tab:results2}}	
    \renewcommand\tabcolsep{0.59cm}
    \renewcommand{\arraystretch}{1.50}
    \centering
    \begin{tabular}{cccc}
        \toprule[1.0pt]
        \toprule[1.0pt]
        \multicolumn{2}{c}{$BB^*$}  &\multicolumn{2}{c}{$B^*B^*$}\\ 
        \hline
        $0(1^{+})$	&$-14.1$        &$0(1^{+})$   &$-13.9$   \\  
        $0(0^{-})$	&$0.8-0.8i$     &$0(1^{-})$   &$7.7-13.7i$   \\
                    &               &$0(2^{-})$   &$0.4-2.5i$   \\
 \bottomrule[1.0pt]
        \bottomrule[1.0pt]
    \end{tabular}
\end{table}

\section{Summary}
Motivated by recent Belle II measurements of the exclusive $e^+e^-\to B^{(*)}\bar{B}^*$ cross section, which reveal the anomalous line shape near the $B\bar{B}^*$ and $B^*\bar{B}^*$ thresholds, we conduct a systematic exploration of the mass spectra of the hidden-bottom and double-bottom molecular tetraquark states up to $P$-wave configurations.

For the $B\bar{B}^*$ and $B^*\bar{B}^*$ systems, we predict a rich spectrum of the hidden-bottom molecular tetraquarks. This includes five bound states: the $B\bar{B}^*$ with $0(1^{++})$ and $0(0^{-+})$, and the $B^*\bar{B}^*$ with $0(2^{++})$, $0(0^{-+})$, and $0(1^{--})$, as well as three resonant states: the $B\bar{B}^*$ with $0(1^{--})$ and $0(2^{--})$, the $B^*\bar{B}^*$ with $0(3^{--})$. Specifically, we found the evidences of two $P$-wave $B\bar{B}^*$ and $B^*\bar{B}^*$ molecules with quantum number $I(J^{PC})=0(1^{--})$ in the $e^+e^-$ annihilation experiments. In parallel, we extend our study to the double-bottom molecular tetraquarks, where the interaction is closely related to that of the hidden-bottom tetraquarks and is determined by applying the $G$-parity rule. In the $0(1^+)$ channel, we identity two bound states for the $BB^*$ and $B^*B^*$ system. Additionally, three $P$-wave resonances are predicted: $BB^*$ with $0(0^-)$, and $B^*B^*$ with $0(1^-)$ and $0(2^-)$.

The discovery of the $X(3872)$, a landmark in exotic hadron physics, has catalyzed two decades of exploration into the non-conventional hadronic configurations. Continuing this momentum, we anticipate a growing spectrum of the hidden- and double-bottom molecular tetraquarks to emerge through collaborative theoretical and experimental efforts. Experimentally, targeted searches near the $B\bar{B}^*$ and $BB^*$ production thresholds at facilities like LHCb and Belle II will be pivotal to confirm the existence of the exotic hadrons in the hidden- and double-bottom molecular tetraquark sector. Key pathways for their discovery mainly include the energy-scan measurement of the $e^+e^- \to B^{(*)}\bar{B}^{(*)}$ cross sections, the reconstruction of displaced vertices in the $pp \to B^{(*)}{B}^{(*)}+X$ processes, and so on. These systems not only expand the zoo of the exotic hadrons but also offer a unique platform to investigate the nonperturbative QCD dynamics. 

\noindent{{\it Note added}}: During the final preparation stages of this manuscript, we noted a contemporaneous work of related research in Ref. \cite{new}. Here, they mainly focused on $P$-wave resonances of the hidden-bottom molecular tetraquark system.

\section*{Acknowledgement}This work is supported by the National Natural Science Foundation of China under Grant Nos. 12247155, 12247101, and 12405097, National Key Research and Development Program of China under Contract No. 2020YFA0406400,  the ‘111 Center’ under Grant No. B20063, the Natural Science Foundation of Gansu Province (No. 22JR5RA389, No. 25JRRA799), the fundamental Research Funds for the Central Universities, the project for top-notch innovative talents of Gansu province, Talent Scientific Fund of Lanzhou University, and the KAKENHI under Grant Nos. 23K03427 and 24K17055.

\end{document}